# Soft X-ray Circular Reflectivity from Ferromagnetic Transition-Metal Films Near the Brewster's Angle: Theoretical and Numerical X-ray Resonant Magnetic Scattering Study

Dae-Eun Jeong and Sang-Koog Kim*

Research Center for Spin Dynamics & Spin-Wave Devices, Seoul National University, Seoul 151-744, Republic of Korea

Nanospintronics Laboratory, Department of Materials Science and Engineering, College of Engineering, Seoul National University, Seoul 151-744, Republic of Korea

**Abstract**

We first report a novel phenomenon that manifests itself in a colossal difference in soft x-ray reflectivity from ferromagnetic transition-metal films between the left- and right-handed circular polarization (LCP and RCP) modes at a resonance near the normal Brewster's angle. Theoretical and numerical studies of soft x-ray resonant magnetic scattering using the circular-polarization-mode basis reveal that this effect arises from a totally destructive interference of photons scattered individually from charge, orbital, and spin degrees of freedom in magnetized thin films that selectively occurs only for one helicity of the opposite circular modes when the required criteria are fulfilled. Across the normal Brewster's angle, the polarization state of scattered soft x rays is continuously variable from the RCP to the LCP mode (or vice versa) through the linear $s$ polarization mode by changing the incidence angle of linearly $p$-polarized x rays at the resonance.

Soft or hard x-ray resonant magnetic scattering (XRMS) measurement techniques have been widely used to investigate charge, orbital, and spin degrees of freedom in multi-component magnetic materials because those techniques offer exceedingly enhanced, element-specific sensitivity to such different orderings at energies close to the absorption edges of a selected element [1-6]. Due to a variety of microscopic interactions between incident photons and each of those orderings, as well as their angular and polarization dependence in the XRMS process, the initial polarization state of incident photons is converted to various polarization states of the scattered photons, which in turn makes it possible to determine element-specific charge, orbital, and spin orderings by analyzing the changed polarization states of the scattered soft x rays and their angular and polarization dependence.

Any arbitrary polarized states of photons can in principle be described in terms of the orthogonal right- and left-handed circular polarization (RCP and LCP) modes (or opposite photon helicities). Since the RCP and LCP modes are not only the basis of an irreducible representation of rotational symmetries in atomic transition processes [7], but are also the eigenmodes of photon beams interacting with the different kinds of orderings in broken symmetries, such circular polarizations are useful in the determination of the fundamental atomic transition spectra in the x-ray resonant region for magnetic materials [8]. For example, material systems with broken symmetries can yield circular-mode-dependent dichroism, such as natural circular dichroism in chiral materials [9, 10], magnetic circular dichroism in magnetic materials [11, 12], and magnetochiral dichroism in parity-nonconserving magnetic materials [13]. Thus, in order to obtain better or deeper insight into not only the interactions of incident photons with different scattering sources of charge, orbital, and spins, but also their polarization and angular dependences, theoretical interpretations of XRMS in the framework of the RCP-and-LCP-modes basis is more fundamental than the linear-polarization-modes basis.

In this Letter, we report the first prediction of differential soft x-ray circular reflectivity from ferromagnetic transition-metal thin films at the resonance energies in a wide angular region across the normal Brewster's angle $\phi_B^n \cong 45°$. It was found that this novel effect arises from a totally destructive interference occurring selectively for either circular polarization mode of photons scattered individually from charge, orbital, and spin degrees of freedom. This can be verified not only by theoretical derivations of the XRMS amplitudes with respect to the circular-mode basis, but also by numerical calculations of the individual intensities of the RCP and LCP components using circular-mode-based magneto-optical Kerr matrices. Also, the continuously variable polarization state of reflected soft x rays from the RCP to the LCP mode (or vice versa) through the linear $s$ polarization mode is observable in a wide angular region across the $\phi_B^n$, and is thus important in that it can be implemented into polarizers or analyzers that enable the control or determination of the polarization states of incoming photons.

First, we derive the XRMS amplitudes with respect to the circular-mode basis by considering a total coherent elastic scattering amplitude in the pure electric dipole ($E1$) transition for the case of $3d$ transition-metal ferromagnetic materials. The amplitude can be expressed by $f_{tot} \cong f_c + f_{xres} + f_m$, where $f_c$, $f_m$, and $f_{xres}$ are the charge, non-resonant magnetic, and resonant scattering contributions, respectively[1, 14, 15]. In general, $f_m$ is noticeably weaker than $f_c$ by a factor of $\hbar\omega/mc^2 \cong 0.002$ at $\hbar\omega = 1\,\text{keV}$, but $f_{xres}$ is comparable to $f_c$ and is much stronger than $f_m$ in the vicinity of the absorption edges [1]. In fact, $f_{xres}$ consists of radial and angular parts. The former is represented by the reduced resonant scattering amplitude, $R'(l_1, j_1; 1; l_2) = \bar{R}(l_1, j_1; 1; l_2) / \left(1 - (\Delta/\Gamma(x-i))^2\right)$ in a fast collision approximation, and the latter by $k$-th rank spin-orbital coupled moments, $M^{(k)}(l_1, j_1; 1; l_2)$ as well as the polarization tensors, $T^{(k)}(\hat{e}_r^*, k_r; \hat{e}_i, k_i)$. Here, $\hat{e}_{r(i)}$ and $\hat{k}_{r(i)}$ represent the polarization and propagation unit vectors of a reflected (incident) photon beam, respectively.

Since the magnetic linear dichroism term $M^{(2)}$ is negligible in the magnitude of the scattering amplitude for $3d$ transition metals [11, 12, 14-17], only the $M^{(0)}$ and $M^{(1)}$ terms are considered in determining the resonant scattering amplitudes with respect to the circular polarization basis, which is given in matrix form by [15, 18]

$$\begin{pmatrix} f_{xres}^R \\ f_{xres}^L \end{pmatrix} = 4\pi\bar{\lambda} R'(l_1, j_1; 1; l_2) \sum_{k=0}^{2} \sum_{q=-k}^{k} \sqrt{\frac{3}{2k+1}} T_q^{(k)*}(\hat{e}_r^*, k_r; \hat{e}_i, k_i) M_q^{(k)}(l_1, j_1; 1; l_2)$$

$$\cong \frac{3}{2}\bar{\lambda} R'(l_1, j_1; 1; l_2) \left( M^{(0)}(l_1, j_1; 1; l_2) \mathbf{A} - \frac{i}{\sqrt{2}} M_m^{(1)}(l_1, j_1; 1; l_2) \mathbf{B} \right) \begin{pmatrix} N_i^R \\ N_i^L \end{pmatrix},$$

(1)

where $M_m^{(1)} = M^{(1)} \cdot \hat{m}$ with a unit vector of magnetization, $\hat{m} = m_1\hat{u}_1 + m_2\hat{u}_2 + m_3\hat{u}_3$. $N_i^{R(L)}$ implies the complex amplitude factor of the RCP(LCP) mode with respect to the circular polarization basis, $\hat{e}_i = N_i^R \hat{e}_i^R + N_i^L \hat{e}_i^L$ . The matrices of **A** and **B** are given as

$\mathbf{A} = \left(\hat{e}_r^* \cdot \hat{e}_i\right)_{RL} = \mathbf{U}_{RL}^{sp\dagger} \left(\hat{e}_r^* \cdot \hat{e}_i\right)_{sp} \mathbf{U}_{RL}^{sp}$ and $\mathbf{B} = \left(\hat{e}_r^* \times \hat{e}_i\right)_{RL} = \mathbf{U}_{RL}^{sp\dagger} \left(\hat{e}_r^* \times \hat{e}_i\right)_{sp} \mathbf{U}_{RL}^{sp}$ . The unitary

operation in the matrices transforms a linear to circular polarization basis or vice versa, which leads to

$$\mathbf{A} = \begin{pmatrix} \frac{1}{2}(1+\cos 2\phi) & \frac{1}{2}(1-\cos 2\phi) \\ \frac{1}{2}(1-\cos 2\phi) & \frac{1}{2}(1+\cos 2\phi) \end{pmatrix} \text{ and } \mathbf{B} = \begin{pmatrix} -\frac{1}{2}m_1 \sin 2\phi - im_2 \cos\phi & \frac{1}{2}m_1 \sin 2\phi - im_3 \sin\phi \\ \frac{1}{2}m_1 \sin 2\phi + im_3 \sin\phi & -\frac{1}{2}m_1 \sin 2\phi + im_2 \cos\phi \end{pmatrix}, \text{ where}$$

$\phi$ is the angle of incidence from the reflection surface as defined in Fig. 1. These matrices represent the

polarization dependence of the total amplitude in the scattering geometry; in particular, **B** shows the

relation of $\phi$ and $\hat{m}$ for each resonant scattering source. For the longitudinal

$(m_2 = \pm 1, m_1 = m_3 = 0)$ and polar magnetizations $(m_3 = \pm 1, m_1 = m_2 = 0)$, the corresponding **B**

are transformed into $\mathbf{B}_{\pm\mathbf{lon}} = \begin{pmatrix} -im_2 \cos\phi & 0 \\ 0 & im_2 \cos\phi \end{pmatrix}$ and $\mathbf{B}_{\pm\mathbf{pol}} = \begin{pmatrix} 0 & -im_3 \sin\phi \\ im_3 \sin\phi & 0 \end{pmatrix}$. The operation of **B** leads to $\hat{e}_i^{R(L)} \to \hat{e}_r^{R(L)}$ for the longitudinal case and $\hat{e}_i^{R(L)} \to \hat{e}_r^{L(R)}$ for the polar case, originating from the conservation of each helicity in the presence of an axial symmetry with respect to $\hat{m}$ [7].

For the specific case of $E1$ resonant scattering at the $L_3$ line from a $3d$ transition metal,

using the relations of $\frac{3}{2}\lambda R' \cong \frac{2}{5}(e/\lambda)^2 \left\{\left[\left(1-\left(\Delta/\Gamma(x-i)\right)^2\right)^{-1} \left|\langle 3d|r|2p_{3/2}\rangle\right|^2\right]/[\varepsilon_l - \varepsilon_b - \hbar\omega - i\Gamma/2]\right\} \equiv \frac{9}{2}iF$ ,

$M^{(0)} \cong \frac{2}{9}\left[N_h(2) - 2\Delta S_m/\Gamma(x-i)\right] \equiv \frac{2}{9}\bar{M}^{(0)}$, $M_m^{(1)} \cong -\frac{1}{9\sqrt{2}}\left[S_m + 3L_m - \Delta N_h(2)/2\Gamma(x-i)\right] \equiv -\frac{1}{9\sqrt{2}}\bar{M}^{(1)}$, the total scattering

amplitudes of the RCP and LCP modes are finally obtained as

$$\begin{pmatrix} f_{tot}^R \\ f_{tot}^L \end{pmatrix} = \left(\left(-r_0 F_c(\vec{K}) + iF\bar{M}^{(0)}\right)\mathbf{A} - \frac{F\bar{M}^{(1)}}{4}\mathbf{B}\right)\begin{pmatrix} N_i^R \\ N_i^L \end{pmatrix}. \tag{2}$$

Eq. (2) implies that each of the charge $r_0 F_c(\vec{K})$, resonant non-magnetic $F\bar{M}^{(0)}$, and resonant magnetic $F\overline{M}^{(1)}$ scattering terms can contribute to $f_{tot}^{R}$ and $f_{tot}^{L}$, by way of the correspondence to the scattering geometry factors represented by the matrices of **A** and **B**. To obtain the individual values of $f_{tot}^{R}$ and $f_{tot}^{L}$, one should consider the individual phase factors of the amplitudes of photons scattered from different scattering sources for each circular mode.

The relations between the phases of photons scattered from the three scattering sources and incident photons for the RCP and LCP modes are described in Table I for the case of an incident $p$ polarization and the longitudinal magnetization of $\hat{m} = +\hat{u}_2$. The phase factors of $+1$, $-1$, $+i$, and $-i$ represent the phase changes in the reflected photons with respect to those of the incident photons, which correspond to $-\pi/2$, $\pi/2$, $0$, and $\pi$, respectively, for $\hat{e}_i^p = -\frac{i}{\sqrt{2}}\hat{e}_i^R + \frac{i}{\sqrt{2}}\hat{e}_i^L$. The asymmetry in these phase factors arises from the phase difference of $\pi$ between the RCP- and LCP-mode components of the incident $p$ polarization. Using the relations shown in Table I, the total scattering amplitudes are given as

$$\begin{pmatrix} f_{tot}^{R} \\ f_{tot}^{L} \end{pmatrix} = \frac{1}{\sqrt{2}} \left( \left( ir_0 F_c(\vec{K}) + F\overline{M}^{(0)} \right) \begin{pmatrix} \cos 2\phi \\ -\cos 2\phi \end{pmatrix} + \frac{F\overline{M}^{(1)}}{4} \begin{pmatrix} \cos\phi \\ \cos\phi \end{pmatrix} \right). \qquad (3)$$

Here, $f_{tot}^{R(L)}$ consists of three different contributions such that $f_{tot}^{R(L)} = f_c^{R(L)} + f_{xres:0}^{R(L)} + f_{xres:1}^{R(L)}$. The phase factors of $f_c^{R(L)}$ and $f_{xres:0}^{R(L)}$ are determined by multiplying the $\cos 2\phi (-\cos 2\phi)$ angular dependence. In the case of $\phi > \pi/4$, the phase factor of $f_c^{R(L)}$ is $-i(+i)$, and that of $f_{xres:0}^{R(L)}$ is of the multiple states of $-1(+1)$ and $+i(-i)$. On the other angular side, $\phi < \pi/4$, the phase factors of $f^R_{c,xres:0}$ and $f^L_{c,xres:0}$ have signs opposite to those for $\phi > \pi/4$. However, the phase factor of $f^{R(L)}_{xres:1}$ does not vary with $\phi$ across $\phi = \pi/4$, but retains the allowed multiple states of $+1$ and $-i$. Furthermore, due to the different angular dependence, that is, $\cos 2\phi$ for both $f^{R(L)}_c$ and $f^{R(L)}_{xres:0}$, and $\cos\phi$ for $f^{R(L)}_{xres:1}$, each value of $f^{R(L)}_c$ and $f^{R(L)}_{xres:0}$ could become comparable in magnitude to $f^{R(L)}_{xres:1}$ at certain angles close to $\phi = \pi/4$. Thus, the asymmetry of the polarization dependence between the RCP and LCP modes gives rise to the differential scattering amplitude, that is, $f^R_{tot} \neq f^L_{tot}$. This inequality can yield differential circular reflectivity for the RCP and LCP modes under specific conditions, where either condition, $f^R_{tot} \neq 0$ and $f^L_{tot} \cong 0$ or $f^L_{tot} \neq 0$ and $f^R_{tot} \cong 0$, can be fulfilled. The illustration of an example of the complete destructive interference for the RCP mode and a non-vanishing interference for the LCP mode are given in Fig. 2. The total amplitude is nearly zero through the vector sum of the imaginary phase terms $f^{R(0)}_{xres:0}$, $f^{R(\pi)}_{xres:1}$, and $f^{R(\pi)}_c$, and the real phase terms $f^{R(\pi/2)}_{xres:0}$ and $f^{R(-\pi/2)}_{xres:1}$ for the RCP mode, but is of a finite value for the counterpart LCP mode. Consequently, we can obtain $f^{R(L)}_{tot} \approx 0$ and $f^{L(R)}_{tot} \neq 0$ at certain circular-mode-dependent Brewster's angles (denoted by $\phi^R_{B,+lon}$ and $\phi^L_{B,+lon}$). The estimate of $\phi^{R(L)}_{B,+lon}$ can be made by solving $f^{R(L)}_{tot} = 0$ at $\phi^{R(L)}_{B,+lon} = \pi/4 + \delta\phi^{R(L)}_{B,+lon}$ for a strong resonant case of $r_0 F_c < F$ in a first-order approximation of $\delta\phi^{R(L)}_{B,+lon}$,

$$\delta\phi^{R(L)}_{B,+lon} \cong +(-)\frac{\sqrt{2}(S_m+3L_m)}{16N_h(2)} - \left(\frac{S_m+3L_m}{16N_h(2)}\right)^2. \tag{4}$$

From Eq. (4), $\Delta\phi_{B,+lon} = \phi^R_{B,+lon} - \phi^L_{B,+lon}$ is given as approximately $\sqrt{2}(S_m+3L_m)/8N_h(2)$. By inserting into Eq. (4) the numerical values of $F_c(\vec{K}) \cong Z = 28$, $F \cong 50 r_0$, $S_m \cong 1.56$, $L_m \cong 0.13$, $N_h(2) \cong 2.5$, $\Gamma \cong 5 eV$, and $\Delta \cong 1.2 eV$ for Co [17, 19, 20], we predicted the numerical values of $\delta\phi_{B,+lon}^{R(L)} = 3.81°(-4.09°)$ and $\Delta\phi_{B,lon} = 7.9°$. This angular difference is somewhat large for 3$d$-transition metals. This allows us to select either the LCP- or RCP-mode component of scattered soft x rays by changing the incidence angle across $\phi_B^n \cong \pi/4$.

To confirm the theoretical XRMS prediction of a colossal difference in soft x-ray circular reflectivity between the opposite photon helicities near $\phi_B^n \cong \pi/4$, we also numerically calculated the intensities of the individual RCP and LCP components as well as the linear $s$ and $p$ components of photons reflected from a model thin film consisting of a Co ($10$ nm) layer on an Si substrate for both cases of $\hat{m} = \pm \hat{u}_2$ with a linearly $p$-polarized incident photon beam at the Co $L_3$ edge. In the calculations, we used the circular-mode-based magneto-optical Kerr matrix [21]. Figure 3(a) shows the angular variations of the individual reflectivities of the $s$- and $p$- and RCP- and LCP-mode components in specular reflection geometry for a demagnetized state of the Co film. In the reflectivity profiles, there exists a $\phi_B^n$ where the reflectivities are extremely low for the linear $p$ polarization, RCP, and LCP modes of the scattered soft x rays, while the reflectivity of the linear $s$ polarization mode is zero in the whole angular range due to no net magnetization. In comparison with the non-magnetized case, contrasting reflectivity profiles between the $s$ and $p$ linear modes, and the RCP and LCP modes are observed near $\phi_B^n$ from the longitudinally magnetized Co film for the incident $p$ polarization, as shown in Fig. 3(b). The numerical values of $\phi_{B,+lon}^{L} = 39.5°$, $\phi_{B,\pm lon}^{p} = 45.2°$, and $\phi_{B,+lon}^{R} = 48.4°$ are observed, as predicted by the circular-mode-based XRMS theory.

At $\phi_{B,\pm lon}^{R(L)}$, the degree of circular polarization represented by the Stokes parameter $S_3$ reaches $+1$ or $-1$, indicating that almost pure circular polarizations can be obtained from the incident $p$ polarization and that the opposite photon helicities can be readily switchable either by changing the incident angle from $\phi_{B,\pm lon}^{L}$ to $\phi_{B,\pm lon}^{R}$ or by reversing the longitudinal magnetization. Also, at $\phi_{B,\pm lon}^{p}$ the degree of linear polarization represented by $S_1$ reaches $+1$, indicating the pure $s$ polarization. The values of $\delta\phi_{B,+lon}^{R(L)} = 3.4°(-5.5°)$ obtained from the numerical calculation are in good agreement with those values of $\delta\phi_{B,+lon}^{R(L)} = 3.81°(-4.09°)$ predicted using Eq. (4). In the above numerical calculation, it was also found that continuously variable polarization states can be obtained by changing $\phi$ across $\phi_B^n$ in the specular reflection. The variation from the RCP to the LCP mode through the linear $s$ polarization mode can be simply obtained by tuning to $\phi_{B,+lon}^{L} = 39.5°$, $\phi_{B,\pm lon}^{p} = 45.2°$, and $\phi_{B,+lon}^{R} = 48.4°$ in a wide angular range, or vice versa for the opposite magnetization orientation, as shown in Fig. 3(d). This novel phenomenon can be implemented into polarizers or analysers that enable the control or determination of the polarization states of incoming photons simply and at very low cost, as multilayer-thin-film linear polarizers [22].

In conclusion, we made a theoretical derivation of the XRMS amplitudes for the individual RCP and LCP modes of soft x rays scattered individually from the charge, orbital and spin degrees of freedom for linearly *p*–polarized incident photons at the resonance.  From this derivation, we found a colossal difference in the soft x-ray circular reflectivity from ferromagnetic transition-metal films over a wide incidence-angle range near the normal Brewster's angle.  This difference originates from a totally destructive interference effect occurring selectively for either the RCP or LCP mode at certain angles across the normal Brewster's angle. The XRMS theory in the framework of the circular-mode basis offers a more fundamental understanding of the polarization and angular-dependent scattering of soft x rays from different scattering sources such as charge, orbital, and spin degrees of freedom in a magnetized material.  Also, the findings on the wide angular difference of the opposite circular-mode-dependent Brewster's angles, as much as $\Delta\phi_{B,lon} = 8.9°$, and on the continuously variable polarization state with slight changes in the incident angle of a linearly *p*-polarized photon beam, might provide practical applications for an optical production (or determination) of the linear and circular components using polarizing elements (or analyzers).

This work was supported by Creative Research Initiatives (ReC-SDSW) of MOST/KOSEF.

References

[1] M. Blume, J. Appl. Phys. **57**, 3615 (1985).

[2] D. Gibbs, G. Grübel, D. R. Harshman, E. D. Isaacs, D. B. McWhan, D. Mills and C. Vettier, Phys. Rev. B **43**, 5663 (1991).

[3] J. B. Kortright, D. D. Awschalom, J. Stöhr, S. D. Bader, Y. U. Idzerda, S. S. P. Parkin, Ivan K. Schuller and H. -C. Siegmann, J. Magn. Magn. Mater. **207**, 7 (1999).

[4] J. B. Kortright and S. -K. Kim, Phys. Rev. B **62**, 12216 (2000).

[5] J. B. Kortright, S. -K. Kim, G. P. Denbeaux, G. Zeltzer, K. Takano, and E. E. Fullerton Phys. Rev. B **64**, 092401 (2001).

[6] G. Srajer, L. H. Lewis, S. D. Bader, A. J. Epstein, C. S. Fadley, E. E. Fullerton, A. Hoffmann, J. B. Kortright, K. M. Krishnan, S. A. Majetich, T. S. Rahman, C. A. Ross, M. B. Salamon, I. K. Schuller, T. C. Schulthess, J. Z. Sun, J. Magn. Magn. Mater. **307**, 1 (2006).

[7] V. B. Berestetskii, E. M. Lifshitz, and L. P. Pitaevskii, *Quantum electrodynamics* (Pergamon Press, Oxford 1982).

[8] G. van der Laan, Phys. Rev. B **57**, 112 (1998).

[9] F. Gel'mukhanov and H. Ågren, Phys. Rep. **312** 87 (1999).

[10] L. Alagna, T. Prosperi, S. Turchini, J. Goulon, A. Rogalev, C. Goulon-Ginet, C. R. Natoli, R. D. Peacock, and B. Stewart, Phy. Rev. Lett. **80**, 4799 (1998).

[11] B. T. Thole, P. Carra, F. Sette, and G. van der Laan, Phys. Rev. Lett. **68**, 1943 (1992).

[12] P. Carra, B. T. Thole, M. Altarelli, and X. Wang, Phys. Rev. Lett, **70**, 694 (1993).

[13] J. Goulon, A. Rogalev, F. Wilhelm, C. Goulon-Ginet, and P. Carra, Phys Rev. Lett. **88**, 237401 (2002).

[14] J. P. Hannon, G. T. Trammell, M. Blume, and D. Gibbs, Phys. Rev. Lett. **61**, 1245 (1988); *ibid* **62**, 2644(E) (1989).

[15] J. Luo, G. T. Trammell, and J. P. Hannon, Phys. Rev. Lett. **71**, 287 (1993).

[16] G. van der Laan, Phys. Rev. Lett. **82**, 640 (1999).

[17] C. Kao, J. B. Hastings, E. D. Johnson, D. P. Siddons, G. C. Smith, and G. A. Prinz, Phys. Rev. Lett. **65**, 373 (1990).

[18] This is a case where the symmetry is $SO_3 \supset SO_2$ with a magnetic ordering and a negligible crystal-field effect.

[19] R. Wu and A. J. Freeman, Phys. Rev. Lett. **73**, 1994 (1994).

[20] S. -K. Kim and J. B. Kortright, Phys. Rev. Lett. **86**, 1347 (2001).

[21] D. -E. Jeong, K. -S. Lee, and S. -K. Kim, Appl. Phys. Lett. **88,** 181109 (2006).

[22] J.B. Kortright, M. Rice, S.-K. Kim, C.C. Walton, T. Warwick J. Magn. Magn. Mater. **191** 79 (1999).

Table Ⅰ. Phase factors of the RCP- and LCP-mode components of photons scattered individually from three different scattering sources, which indicate the phase change of the scattered photons with respect to those of the RCP- and LCP-mode components of a linearly $p$-polarized incident photon beam. In the notation of $\alpha\beta_n^{(\gamma)}$, $\alpha$ and $\beta$ indicate the circular polarization components in the incident and reflected photons, respectively. The $\gamma$ and $n$ variables represent the phase difference and the scattering source, respectively.

| | $\hat{e}_i^R \rightarrow \hat{e}_r^R$ | $\hat{e}_i^L \rightarrow \hat{e}_r^R$ | $\hat{e}_i^R \rightarrow \hat{e}_r^L$ | $\hat{e}_i^L \rightarrow \hat{e}_r^L$ |
|---|---|---|---|---|
| Charge scattering (c) | $+i : RR_c^{(0)}$ | $-i : LR_c^{(\pi)}$ | $+i : RL_c^{(0)}$ | $-i : LL_c^{(\pi)}$ |
| Resonant non-magnetic scattering (xres : 0) | $+1 : RR_{xres:0}^{(-\pi/2)}$ <br> $-i : RR_{xres:0}^{(\pi)}$ | $-1 : LR_{xres:0}^{(\pi/2)}$ <br> $+i : LR_{xres:0}^{(0)}$ | $+1 : RL_{xres:0}^{(-\pi/2)}$ <br> $-i : RL_{xres:0}^{(\pi)}$ | $-1 : LL_{xres:0}^{(\pi/2)}$ <br> $+i : LL_{xres:0}^{(0)}$ |
| Resonant magnetic scattering (xres : 1) | $+1 : RR_{xres:1}^{(-\pi/2)}$ <br> $-i : RR_{xres:1}^{(\pi)}$ | Not allowed | Not allowed | $+1 : LL_{xres:1}^{(-\pi/2)}$ <br> $-i : LL_{xres:1}^{(\pi)}$ |

**Figure captions**

FIG. 1. (color online) Definition and coordinate system used in the text for a specular reflection geometry with the linear $s$- and $p$- as well as RCP- and LCP-mode bases in the incident and reflected photons.

FIG. 2. (color online) Illustration of interferences in each of the RCP- and LCP-mode components of photons scattered individually from the charge $f_c$, resonant non-magnetic $f_{xres:0}$, and resonant magnetic $f_{xres:1}$ scattering sources. The radius of each circle represents the magnitude of the corresponding scattering amplitudes. The direction of the arrow inside each circle indicates the corresponding phase factor, as noted in the inset, which is also described by $\gamma$ in $f^{R,L(\gamma)}$.

FIG. 3. (color online) Calculations of the intensities of the linear $s$, $p$ polarization and of the RCP- and LCP-mode components at the Co $L_3$ edge for incident $p$-polarized x rays for (a) the demagnetization state of Co and (b) the longitudinally magnetized states of $\hat{m} = \pm\hat{u}_2$. The superscripts and subscripts in $I^{s,p,R,L}_{+,-}$ denote the corresponding polarization component in the scattered photons and either state of $\hat{m} = \pm\hat{u}_2$, respectively. (c) Calculations of the Stokes parameters, $S_1$, $S_2$ and $S_3$ as a function of $\phi$ for $\hat{m} = +\hat{u}_2$. The relations of the Stokes parameters and the degree of linear $P_L$ or circular $P_C$ polarization are expressed by $S_3 = P_C$ and $S_1^2 + S_2^2 = P_L^2$. (d) Continuously variable polarization states at the indicated angles in a wide angular region across the normal Brewster's angle for $\hat{m} = +\hat{u}_2$.

**Figure 1.**

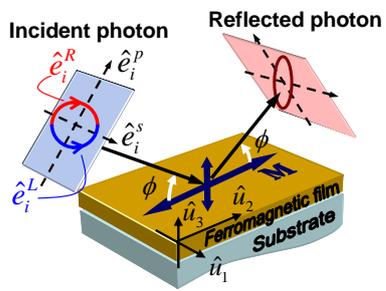

**Figure 2.**

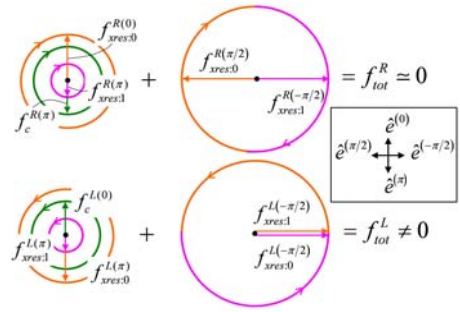

**Figure 3.**

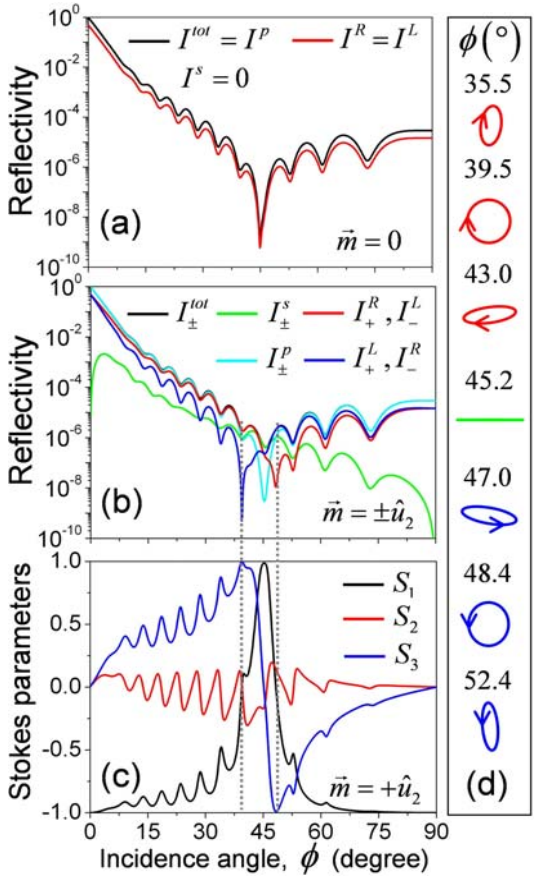